\documentclass[preprintnumbers,10pt,nofootinbib]{revtex4}
\pdfoutput=1

\usepackage{amsmath,latexsym,amssymb,amsfonts}
\usepackage[pdftex]{color,graphicx}
\usepackage{bm}

\begin{document}


\title{\textbf{Two interpretations of thin-shell instantons}}

\author{
\textsc{Pisin Chen}$^{a,b,c,d}$\footnote{{\tt pisinchen{}@{}phys.ntu.edu.tw}}, \textsc{Yao-Chieh Hu}$^{a,b,c}$\footnote{{\tt r04244003{}@{}ntu.edu.tw}} and \textsc{Dong-han Yeom}$^{a}$\footnote{{\tt innocent.yeom{}@{}gmail.com}}
}

\affiliation{
$^{a}$\small{Leung Center for Cosmology and Particle Astrophysics, National Taiwan University, Taipei 10617, Taiwan}\\
$^{b}$\small{Department of Physics, National Taiwan University, Taipei 10617, Taiwan}\\
$^{c}$\small{Graduate Institute of Astrophysics, National Taiwan University, Taipei 10617, Taiwan}\\
$^{d}$\small{Kavli Institute for Particle Astrophysics and Cosmology, SLAC National Accelerator Laboratory, Stanford University, Stanford, California 94305, USA}\\
}

\begin{abstract}
For $O(4)$-symmetric instantons, there are two complementary interpretations for their analytic continuations. One is the nothing-to-something interpretation, where the initial and final hypersurfaces are disconnected by Euclidean manifolds. The other is the something-to-something interpretation, introduced by Brown and Weinberg, where the initial and final hypersurfaces are connected by the Euclidean manifold. These interpretations have their own pros and cons and hence they are complementary. In this paper, we consider analytic continuations of thin-shell instantons that have less symmetry, i.e., the spherical symmetry. When we consider the Farhi-Guth-Guven/Fischler-Morgan-Polchinski tunneling, the something-to-something interpretation has been used in the usual literature. On the other hand, we can apply the nothing-to-something interpretation with some limited conditions. We argue that for both interpretations, we can give the consistent decay rate. As we apply and interpret what follows the nothing-to-something interpretation, a stationary black hole can emit an expanding shell that results in a spacetime without a singularity or event horizon.
\end{abstract}

\maketitle

\newpage

\tableofcontents

\section{Introduction}

One of the main goals of modern theoretical physics is to develop a consistent theory of quantum gravity. This consistent quantum theory of gravity will help us to understand two important and interesting physical problems; one is the information loss problem of a black hole \cite{Hawking:1976ra} and the other is the initial singularity problem of the universe \cite{Hawking:1969sw}. Even though there is no consensus on quantum gravity yet, we already briefly know that these problems should be fairly discussed by quantizing the spacetime, i.e., by investigating the wave function of the universe via the Wheeler-DeWitt equation \cite{DeWitt:1967yk}.

The Euclidean path integral approach gives good wisdom for these two problems \cite{Hartle:1983ai}. Although the Euclidean path integral is not a complete approach in the sense that the path integral is not bounded from below, this Euclidean path integral is at least a good approximation of the ground state wave function. This wave function can be well approximated by solving on-shell solutions, so-called instantons. These instanton solutions present approximate but very important contributions as non-perturbative effects; and all of these results will probably not be largely changed even when we eventually know a consistent theory of quantum gravity.

In this perspective, we focus on classical and quantum behaviors of thin-shell bubbles in Einstein gravity. By using the thin-shell approximation \cite{Israel:1966rt}, we can investigate not only the $O(4)$ symmetry, but also the spherical symmetry. This means that now we can deal with non-perturbative effects of black holes and hence this can be related to the information loss problem (for further review, see \cite{Chen:2014jwq}). As Maldacena \cite{Maldacena:2001kr} and Hawking \cite{Hawking:2005kf} have pointed out, non-perturbative effects of a black hole will shed some light on the information loss problem \cite{Sasaki:2014spa,Lee:2015rwa,Chen:2015lbp}.

More specifically, in this paper, we are interested in the interpretation of instantons. By interpretation, we mean the way to analytically continue instantons to Lorentzian signatures, while an instanton itself is a solution in the Euclidean signatures. As we discuss in the following sections, in the $O(4)$-symmetric instantons \cite{Coleman:1980aw,Hawking:1981fz}, there are two competitive interpretations; one is that the initial and final hypersurfaces are separated by instantons and the other is that the initial and final hypersurfaces are connected by instantons \cite{Brown:2007sd}. The former is a more mathematically complete interpretation, while the latter is a more natural generalization from the interpretation of the Minkowski case. In this paper, we regard the two interpretations as being complementary to each other. In addition, we generalize this complementary interpretation to thin-shell instantons that have less symmetry than the $O(4)$ symmetry. This helps us to see the same instanton with a different point of view; and we may find interesting solutions that will be helpful in understanding the information loss problem of black holes.

This paper is organized as follows. In SEC.~\ref{sec:ana}, we summarize two complementary interpretations for $O(4)$-symmetric instantons. In SEC.~\ref{sec:dyn}, we generalize these interpretations to thin-shell instantons. We discuss that for some limited cases, we can interpret thin-shell instantons such that the final hypersurface is disconnected from the initial hypersurface. In this case, we can further argue that a stationary black hole can disappear into a trivial geometry without a singularity or an event horizon by emitting an out-going shell. This sheds some light on the information loss problem. Finally, in SEC.~\ref{sec:dis}, we summarize our results and discuss possible future issues. In this paper, we use the convention that $c = \hbar = G = 1$.

\section{\label{sec:ana}Analytic continuation of de Sitter space: $O(4)$ symmetry}

\subsection{Coordinates of Euclidean de Sitter space}

\begin{figure}
\begin{center}
\includegraphics[scale=0.6]{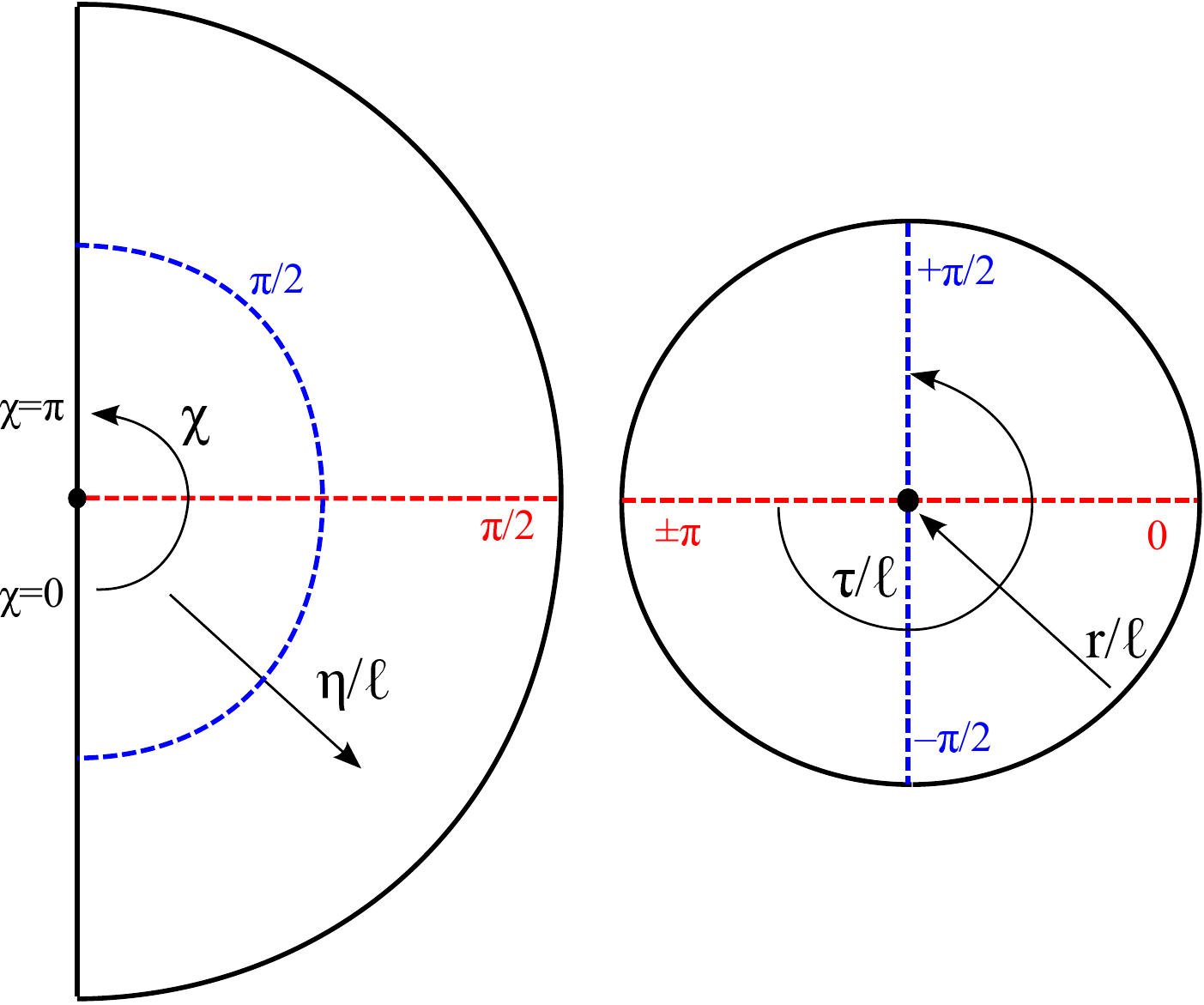}
\caption{\label{fig:bounce_slice}Coordinate patch of the Euclidean de Sitter space with the $\eta$-$\chi$ coordinate (left: $\eta$ is the radial direction and $\chi$ is the angular direction) and $\tau$-$r$ coordinate (right: $\tau$ is the angular direction and $r$ is the radial direction). For inhomogeneous tunneling, we paste Euclidean and Lorentzian manifolds at the red dotted line. For homogeneous tunneling, we paste the manifolds at the blue dotted curve.}
\end{center}
\end{figure}

We can describe a Euclidean de Sitter space with the cosmological constant $\Lambda = 1/\ell^{2}$ by two well-known coordinates \cite{Brown:2007sd}: either the time-dependent form (left of FIG.~\ref{fig:bounce_slice})
\begin{eqnarray}
ds^{2} = d\eta^{2} + \rho^{2}(\eta) \left(d\chi^{2} + \sin^{2}\chi d\Omega^{2}\right),
\end{eqnarray}
where
\begin{eqnarray}
\rho = \ell \sin \frac{\eta}{\ell}
\end{eqnarray}
and variables cover
\begin{eqnarray}
&&0 \leq \frac{\eta}{\ell} \leq \pi, \\
&&0 \leq \chi \leq \pi
\end{eqnarray}
or the time-independent form (right of FIG.~\ref{fig:bounce_slice})
\begin{eqnarray}
ds^{2} = \left( 1 - \frac{r^{2}}{\ell^{2}} \right)d\tau^{2} + \left( 1 - \frac{r^{2}}{\ell^{2}} \right)^{-1} dr^{2} + r^{2} d\Omega^{2},
\end{eqnarray}
where
\begin{eqnarray}
&&0 \leq r \leq \ell, \\
&&-\pi \leq \frac{\tau}{\ell} \leq \pi.
\end{eqnarray}
These two coordinates are connected by the following relations:
\begin{eqnarray}
r &=& \ell \sin \frac{\eta}{\ell} \sin \chi,\\
\tan \frac{\tau}{\ell} &=& \tan \frac{\eta}{\ell} \cos \chi.
\end{eqnarray}

\subsection{Analytic continuations}

When we interpret the inhomogeneous tunneling \cite{Coleman:1980aw}, we do the Wick-rotation along the $\chi = \pi/2$ slice (red dashed line in the left of FIG.~\ref{fig:bounce_slice}) \cite{Hawking:1998bn}, where this hypersurface satisfies the conditions
\begin{eqnarray}
r &=& \rho(\eta),\\
\frac{\tau}{\ell} &=& 0, \;\; \pm \pi
\end{eqnarray}
and hence is equivalent with the red dashed line in the right of FIG.~\ref{fig:bounce_slice}.

On the other hand, when we interpret the homogeneous tunneling \cite{Hawking:1981fz,Lee:2012qv}, we do the Wick-rotation along the $\eta/\ell = \pi/2$ slice (blue dashed curve in the left of FIG.~\ref{fig:bounce_slice}), where it satisfies
\begin{eqnarray}
r &=& \ell \sin \chi,\\
\frac{\tau}{\ell} &=& \pm \frac{\pi}{2},
\end{eqnarray}
and hence is equivalent with the blue dashed curve in the right of FIG.~\ref{fig:bounce_slice}.

\begin{figure}
\begin{center}
\includegraphics[scale=0.6]{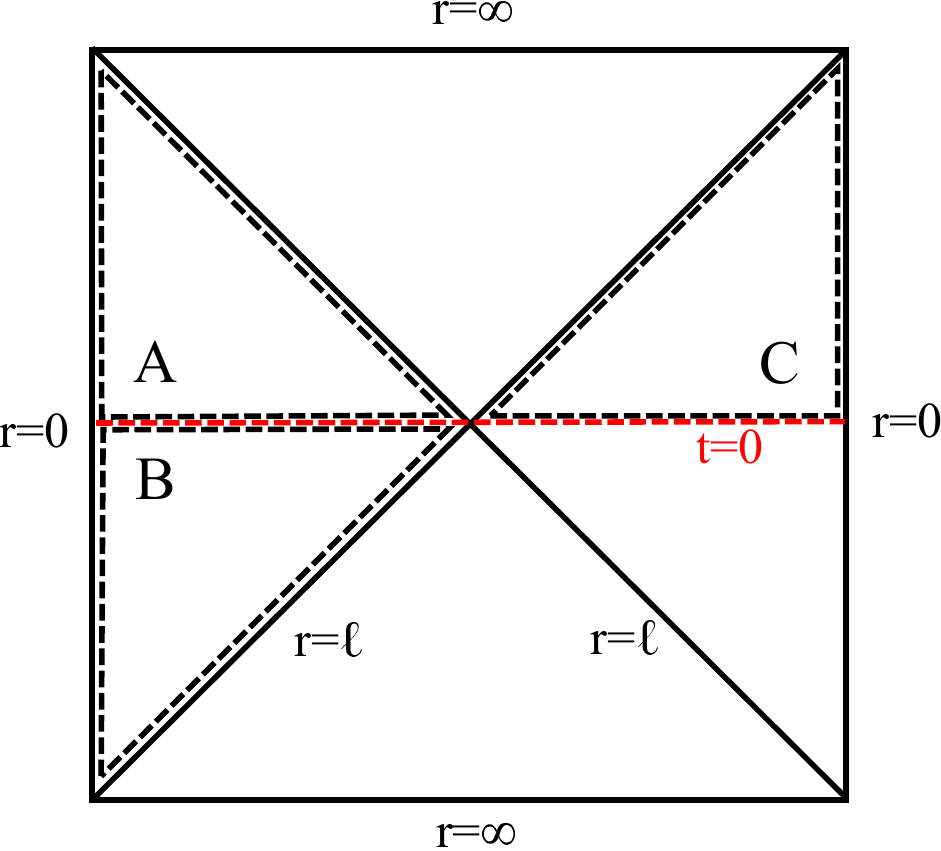}
\caption{\label{fig:bounce_lorentz}Penrose diagram of static Lorentzian de Sitter space. $A$, $B$, and $C$ are pieces of Lorentzian de Sitter space that are analytically continued by Euclidean manifolds (FIG.~\ref{fig:analytic_continuation}).}
\end{center}
\end{figure}

In this paper, we are interested in the inhomogeneous tunneling. Then we can identify the red dashed line of FIG.~\ref{fig:bounce_slice} with the red dashed line of FIG.~\ref{fig:bounce_lorentz}. In this regard, there are two ways to interpret.

\begin{figure}
\begin{center}
\includegraphics[scale=0.6]{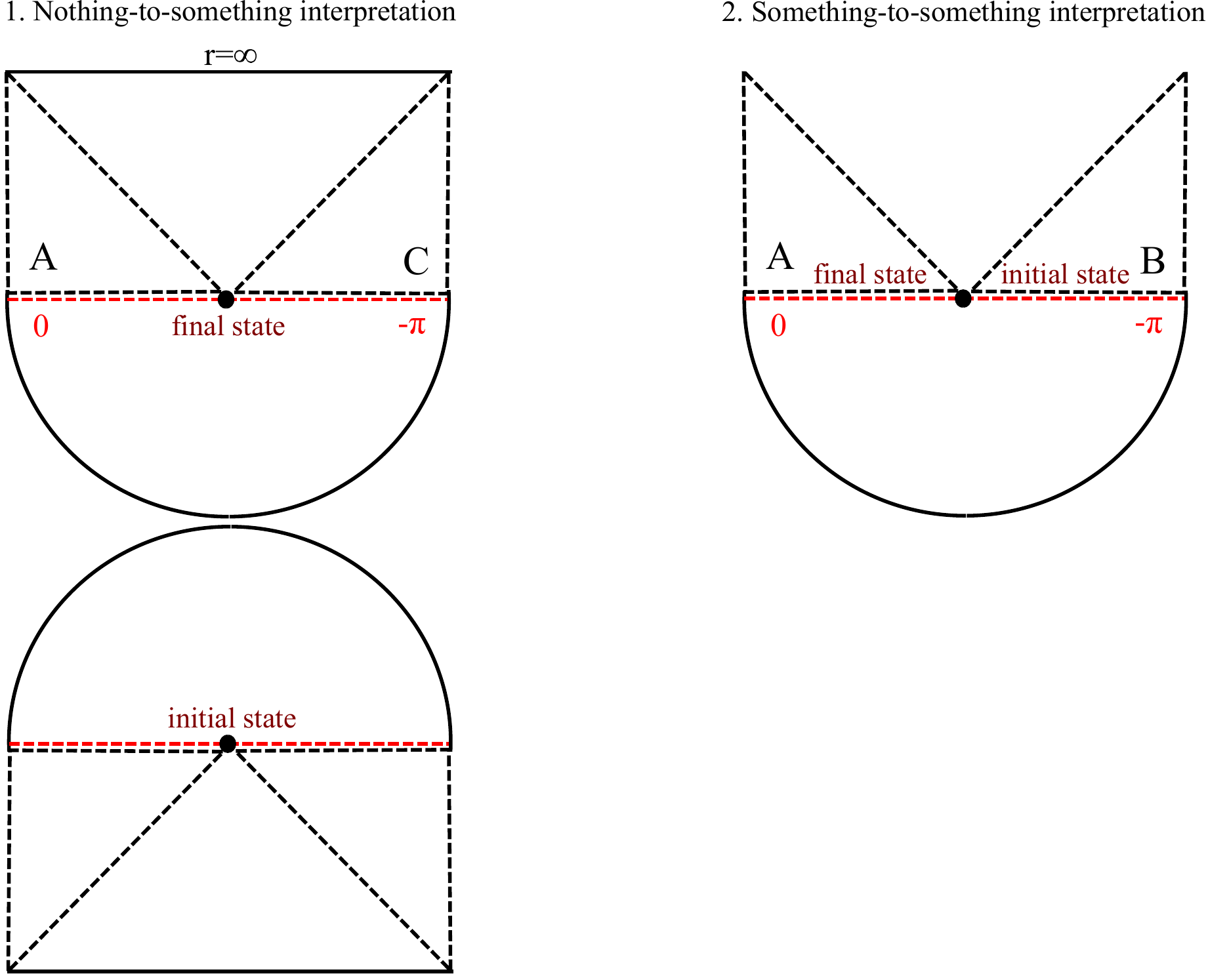}
\caption{\label{fig:analytic_continuation}There are two ways to paste Euclidean and Lorentzian de Sitter space. Left: we paste all of the complete manifold after the $t=0$ slice. Then we interpret that a Lorentzian de Sitter space is created from nothing. Right: we paste slices of $A$ and $B$. Then $B$ is the initial state and $A$ is the final state. Two states are connected by the instanton.}
\end{center}
\end{figure}

\paragraph{Nothing-to-something interpretation}

One way is to paste the entire hypersurface from the Euclidean manifold to the Lorentzian manifold (left of FIG.~\ref{fig:analytic_continuation}). Then the initial state and final state include both the left and right side of the Lorentzian causal patches ($A$ and $C$ of FIG.~\ref{fig:bounce_lorentz}). The Lorentzian-Euclidean combined manifold for the initial state is disconnected to that of the final state. In this sense, one Lorentzian-Euclidean combined manifold can be interpreted such that the manifold is created from nothing\footnote{However, this does not mean that there is no initial hypersurface. There can be an initial hypersurface, but the initial hypersurface and final hypersurface are \textit{disconnected} by the instanton.}. In this paper, we call this the nothing-to-something interpretation.

\paragraph{Something-to-something interpretation}

Brown and Weinberg suggested an alternative interpretation~\cite{Brown:2007sd}. Mathematically it is possible to paste $B$ to the right part of the Euclidean manifold ($\tau/\ell = - \pi$) and paste $A$ to the left part of the Euclidean manifold ($\tau/\ell = 0$): the right of FIG.~\ref{fig:analytic_continuation}. Then we interpret that $B$ is the initial state and $A$ is the final state. If the instanton solution is non-trivial for the region $B$ \cite{Lee:2012qv,Hackworth:2004xb}, one needs to interpret that a thermal excitation created the non-trivial field combination on $B$.

\subsection{Pros and cons: motivation of this paper}

Both of previous approaches give the same decay rate. Therefore, in terms of the calculations, we cannot distinguish which is true. However, these two different interpretations may have pros and cons. We illustrate these as follows.
\begin{itemize}
\item[--] \textit{Nothing-to-something interpretation, pros}: This is mathematically natural. The Euclidean-Lorentzian joined manifold is entirely smooth and maximally extended.
\item[--] \textit{Nothing-to-something interpretation, cons}: This interpretation needs to cover beyond the Hubble radius $r = \ell$, which is outside one's causal patch where may be unphysical.
\item[--] \textit{Something-to-something interpretation, pros}: Everything happens inside one's causal patch. The initial state and final state are connected by the Euclidean manifold, where this corresponds well with the case without gravity.
\item[--] \textit{Something-to-something interpretation, cons}: The Euclidean-Lorentzian joined manifold cannot be maximally extended. If the scalar field is non-trivial beyond the Hubble radius, then it needs to rely on the thermal excitation, which is quite subtle. In addition, this interpretation cannot be applied for anti-de Sitter spaces.
\end{itemize}

In this paper, we regard these two interpretations as complementary approaches. We apply beyond the $O(4)$ symmetry such as to spherical symmetry and see that there is a possibility to interpret using both ways.


\section{\label{sec:dyn}Dynamics and tunneling of thin-shell bubbles}

\subsection{Equation of motion}

We investigate the Einstein gravity with a scalar field,
\begin{eqnarray}
S = \int_{\mathcal{M}} \sqrt{-g} d^{4}x \left[ \frac{\mathcal{R}}{16 \pi} - \frac{1}{2} \nabla^{\mu} \phi \nabla_{\mu} \phi - U(\phi) \right] + \int_{\partial \mathcal{M}} \sqrt{-h} d^{3}x \left[ \frac{\mathcal{K} - \mathcal{K}_{0}}{8 \pi} \right],
\end{eqnarray}
where $g_{\mu\nu}$ is the metric, $\mathcal{R}$ is the Ricci scalar, $\phi$ is a scalar field, $U(\phi)$ is a potential of the scalar field, $\mathcal{K}$ is the Gibbons-Hawking boundary term \cite{Gibbons:1976ue} at a hypersurface $h$ (which is the boundary $\partial \mathcal{M}$ of the entire manifold $\mathcal{M}$), and $\mathcal{K}_{0}$ is the Gibbons-Hawking boundary term of the Minkowski metric.

As a toy model, we consider a true vacuum bubble in the Schwarzschild background with the thin-shell approximation. That is, we use the following metrics for inside and outside the shell:
\begin{eqnarray}
ds^{2} = - f_{\pm}(R) dT^{2} + \frac{1}{f_{\pm}(R)} dR^{2} + R^{2} d\Omega^{2},
\end{eqnarray}
where $+$ denotes outside the shell, $-$ denotes inside the shell, and the metrics satisfy
\begin{eqnarray}
f_{\pm} = 1 - \frac{2M_{\pm}}{R} + \frac{R^{2}}{L_{\pm}^{2}},
\end{eqnarray}
where $M_{-} = 0$, $L_{+} = \infty$, $M_{+} = M > 0$, and $M$ and $L_{-}$ are free parameters. In addition, the shell is located at $r(t)$ with the induced metric
\begin{eqnarray}
ds_{\mathrm{shell}}^{2} = - dt^{2} + r^{2}(t) d\Omega^{2}. 
\end{eqnarray}

According to the well-known Israel's junction equation \cite{Israel:1966rt}, we can derive the equation of motion:
\begin{eqnarray}
\epsilon_{-} \sqrt{\dot{r}^{2} + f_{-}} - \epsilon_{+} \sqrt{\dot{r}^{2} + f_{+}} = 4\pi r \sigma,
\end{eqnarray}
where $\sigma$ is a constant tension parameter and $\epsilon_{\pm} = \pm 1$ denotes the outward normal directions outside and inside the shell, respectively. Here, $\epsilon_{\pm}$ should be proportional to the extrinsic curvatures $\beta_{\pm}$, where
\begin{eqnarray}
\beta_{\pm} \equiv \frac{f_{-} - f_{+} \mp 16 \pi^{2} \sigma^{2} r^{2}}{8\pi \sigma r}.
\end{eqnarray}
In our setting with $M_{-} = 0$ and $L_{+} = \infty$, we can easily check that $\beta_{-} > 0$. Regarding $\beta_{+}$, in the large $r$ limit, if $4 \pi \sigma L_{-} < 1$, then $\beta_{+} > 0$ is also satisfied. In this paper, we consider these limits so that every tunneling happens within the right patch of the Penrose diagram of the Schwarzschild solution (see SEC.~\ref{sec:speint}).

\begin{figure}
\begin{center}
\includegraphics[scale=0.6]{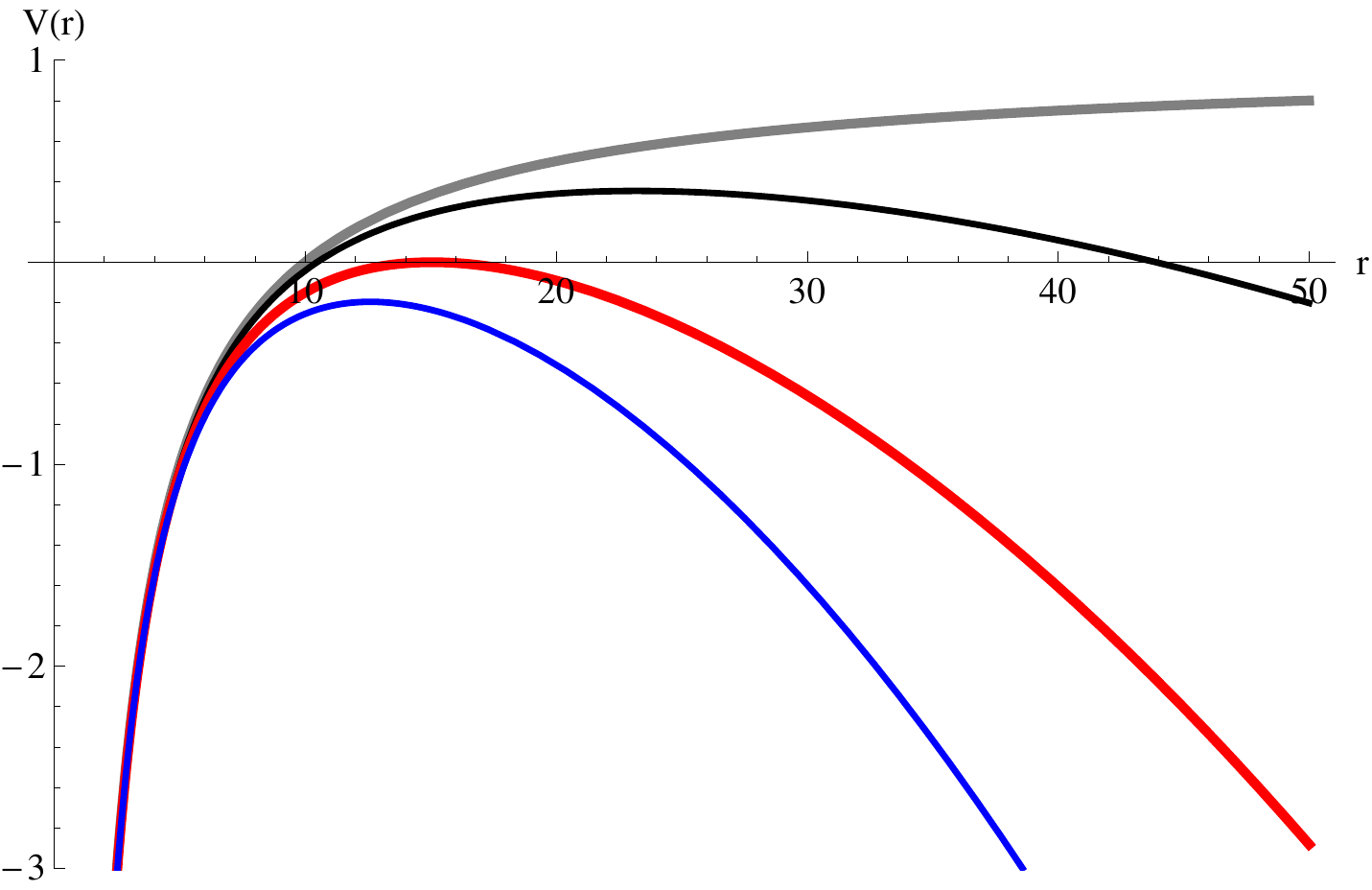}
\caption{\label{fig:effpotential}An example of the effective potential $V(r)$ for $M = 5$ and $L_{-} = 0.1$, varying the tension $\sigma_{0} = 1/4\pi L_{-}$ (gray: the extreme case in which $\Delta \tau = \infty$), $\sigma_{1} = 0.998 \times \sigma_{0}$ (black), $\sigma_{2} \simeq 0.99617 \times \sigma_{0}$ (red: when two zeros are degenerate), and $\sigma_{3} = 0.995 \times \sigma_{0}$ (blue: when there is no zero).}
\end{center}
\end{figure}

Finally, the junction equation can be simplified as
\begin{eqnarray}
\dot{r}^{2} + V(r) &=& 0,\\
V(r) &\equiv& f_{+} - \frac{(f_{-} - f_{+} - 16 \pi^{2}\sigma^{2} r^{2})^{2}}{64 \pi^{2} \sigma^{2} r^{2}}.
\end{eqnarray}
Note that $V(r)$ always goes to $-\infty$ for the $r \rightarrow 0$ limit or the $r \rightarrow \infty$ limit if $4 \pi \sigma L_{-} < 1$ (FIG.~\ref{fig:effpotential}).

\subsection{Farhi-Guth-Guven/Fischler-Morgan-Polchinski tunneling}

We especially consider the case when $V(r) = 0$ has two solutions, say $r_{1} < r_{2}$. If $r_{1} \leq r \leq r_{2}$, then the shell is classically forbidden, while quantum mechanically we can consider a tunneling between $r_{1}$ and $r_{2}$, or vice versa \cite{Farhi:1989yr,Fischler:1989se}. Originally, Farhi-Guth-Guven \cite{Farhi:1989yr} and Fischler-Morgan-Polchinski \cite{Fischler:1989se} considered tunneling of a \textit{false} vacuum bubble, while in this paper we will consider a \textit{true} vacuum bubble (hence, inside is anti-de Sitter space). This is just a technical reason for choosing positive extrinsic curvatures. However, we can adopt the same techniques of Farhi-Guth-Guven and Fischler-Morgan-Polchinski. In this sense, we name this tunneling process Farhi-Guth-Guven/Fischler-Morgan-Polchinski tunneling.

\begin{figure}
\begin{center}
\includegraphics[scale=0.5]{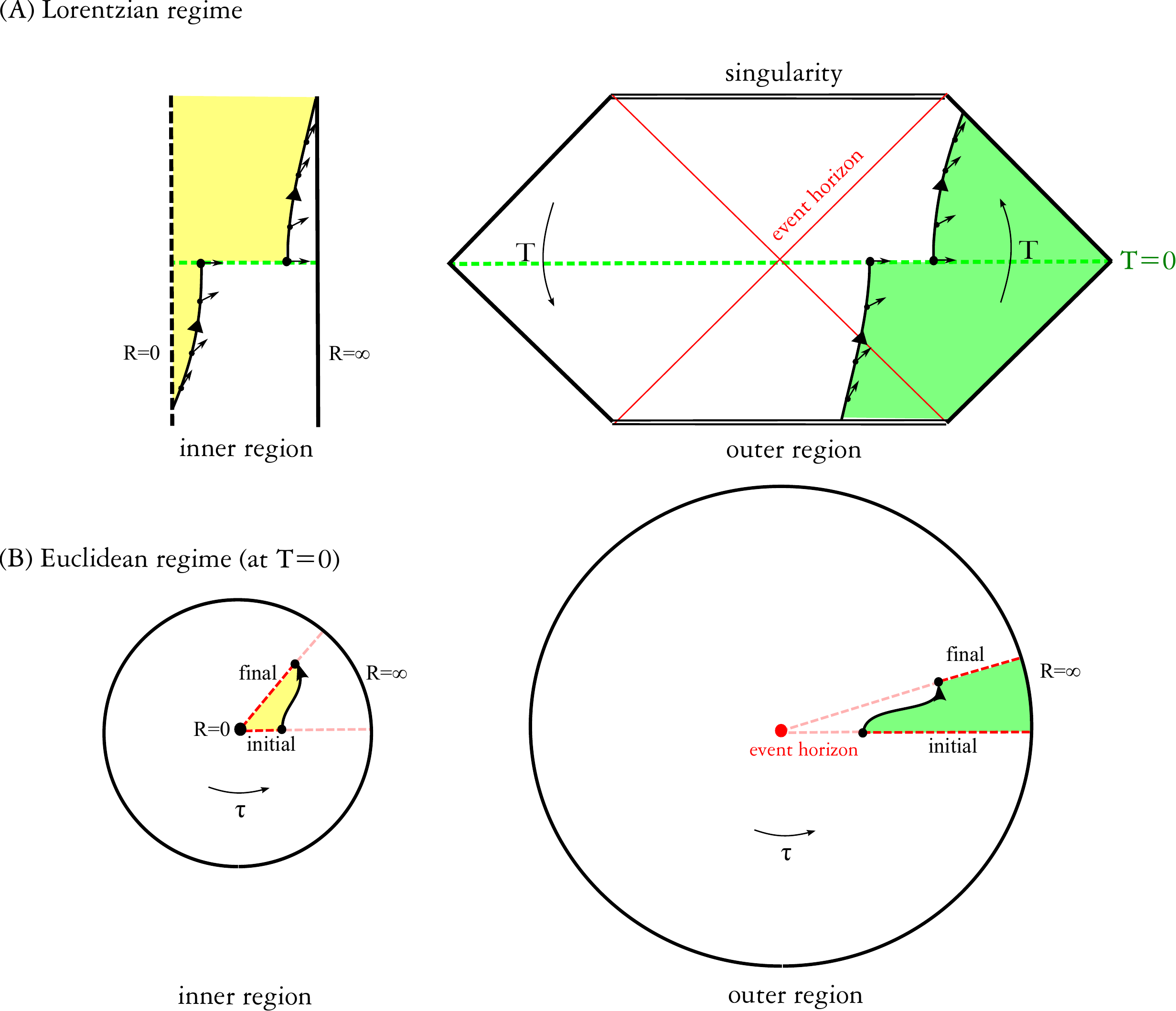}
\caption{\label{fig:FGG_TV}Farhi-Guth-Guven tunneling for a true vacuum bubble case. The left of (B) is a periodically identified anti-de Sitter space at $R = 0$ and hence $R = 0$ is regular for any period, while the right of (B) is a Euclidean Schwarzschild with the time period $\Delta \tau_{+} = 8\pi M$; if the time period is different from $8\pi M$, then there appears a cusp singularity on the red dot (event horizon).}
\end{center}
\end{figure}

\subsubsection{Usual interpretation: something-to-something}

FIG.~\ref{fig:FGG_TV} is the usual and traditional interpretation of Farhi-Guth-Guven tunneling \cite{Farhi:1989yr}. The upper diagram is the shell dynamics in the Lorentzian signatures. The left of (A) is the anti-de Sitter space and the right of (A) is the Schwarzschild space. Initially, the shell starts from $r = 0$ and expands up to its maximum radius $r_{1}$. After the tunneling, the shell reaches $r_{2}$ and expands toward infinity. The lower diagram is the shell dynamics in the Euclidean signatures. The shell moves from $r_{1}$ to $r_{2}$. As we identify the initial and final surface, the Euclidean manifold connects from the initial to the final surface, and hence this is the something-to-something interpretation. Note that the left of (B) is periodically identified as Euclidean anti-de Sitter and hence there is no cusp singularity at $R = 0$ with any period. On the other hand, for the right of (B), in order to avoid the cusp singularity at the event horizon, we need to use the exact Euclidean time period $8 \pi M$.

\begin{figure}
\begin{center}
\includegraphics[scale=0.5]{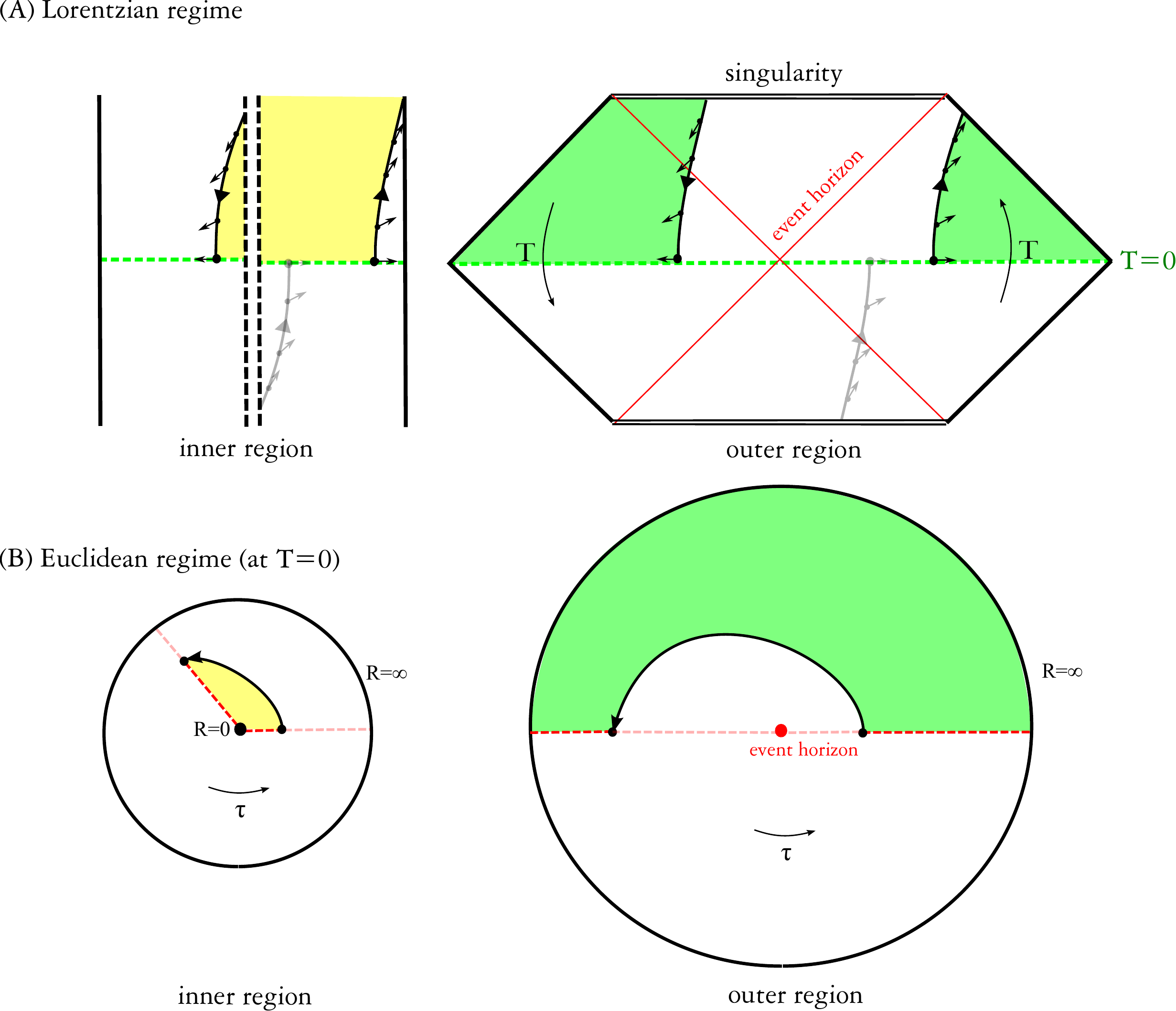}
\caption{\label{fig:FGG_TV2}If the Euclidean time period of the thin-shell corresponds to that of the event horizon of the Euclidean Schwarzschild, then one can do both interpretations: something-to-something (gray) or nothing-to-something (black). Here, (A) is the shell dynamics for the Lorentzian regime and (B) is for the Euclidean regime.}
\end{center}
\end{figure}

\subsubsection{\label{sec:speint}Special interpretation: nothing-to-something}

\begin{figure}
\begin{center}
\includegraphics[scale=0.5]{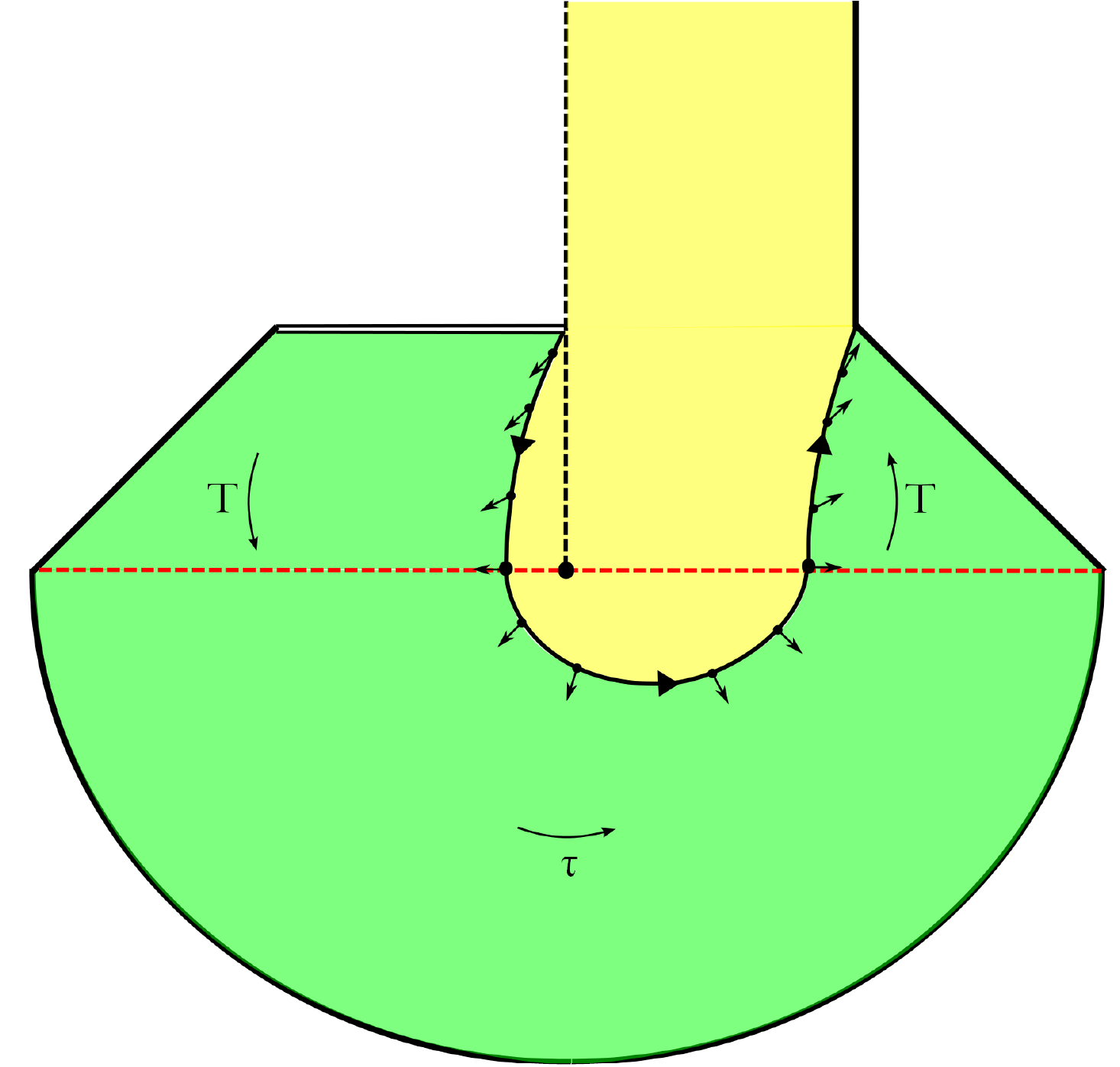}
\caption{\label{fig:ads_single}One can eventually interpret following the nothing-to-something interpretation.}
\end{center}
\end{figure}

If the time period of the shell $\Delta \tau$ (the Euclidean time for the process in which the shell starts from $r_{1}$, reaches $r_{2}$, and returns back to $r_{1}$) corresponds to the time period of the background $8\pi M$, i.e., after considering the correct redshift, if the condition is satisfied that
\begin{eqnarray}\label{eq:redshi}
8\pi M = \Delta \tau_{+} \equiv \int_{0}^{\Delta \tau} d\tau \frac{\sqrt{f_{+}-V(r)}}{f_{+}},
\end{eqnarray}
then one can apply not only the something-to-something interpretation, but also the nothing-to-something interpretation. If $\Delta \tau_{+}$ is not the same as $8\pi M$, then in the action integration, the boundary term at infinity\footnote{In interpreting the nothing-to-something interpretation, the importance of the boundary term at infinity was overlooked in Gregory-Moss-Withers \cite{Gregory:2013hja}, since the authors were interested in the de Sitter background. On the other hand, in the Minkowski background, we surely need to include the boundary term at infinity \cite{Gibbons:1976ue} and in order to cancel out this term between the initial and final surfaces, we need to restrict $\Delta \tau_{+}$.} of the background ($= 4 \pi M^{2}$) cannot be canceled to that of the solution ($= \Delta \tau_{+} M/2$). By fixing the solution period as $8\pi M$, one may need to worry whether there appears a cusp singularity in the inside geometry or not. If $M_{-} > 0$, then unless $M_{-} = M_{+}$, there is a cusp singularity at the horizon. However, if $M_{-} = 0$, then one can periodically identify with an arbitrary period at $R = 0$; hence, in our examples, there is no problem. (When there appears a cusp singularity, we comment on this later: see SEC.~\ref{sec:comm}.) FIG.~\ref{fig:FGG_TV2} is the new interpretation. By pasting inside and outside geometry, we obtain FIG.~\ref{fig:ads_single}. One can notice that this Euclidean-Lorentzian joined manifold is disconnected from the initial Schwarzschild black hole and hence this is indeed the nothing-to-something interpretation. As a simple generalization, one can further find more general cases: $\Delta \tau_{+} \times N = 8 \pi M$, where $N$ is a natural number.

\begin{figure}
\begin{center}
\includegraphics[scale=0.5]{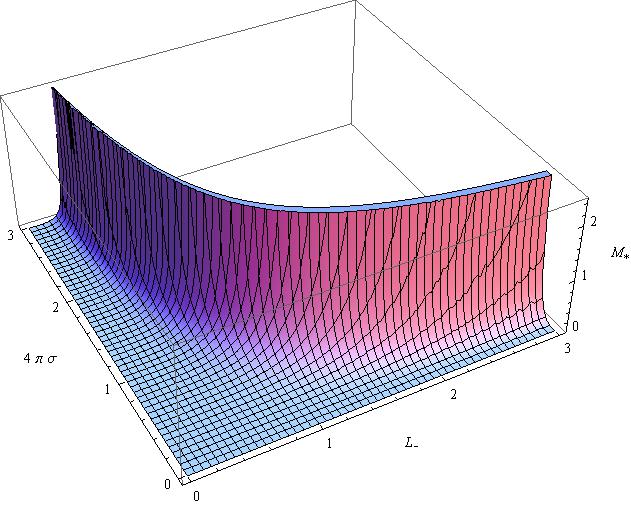}
\caption{\label{fig:range}$M_{*}$ as a function of $L_{-}$ and $4 \pi \sigma$, where we restricted the region by $4 \pi \sigma L_{-} \leq 1$.}
\end{center}
\end{figure}

\begin{figure}
\begin{center}
\includegraphics[scale=0.7]{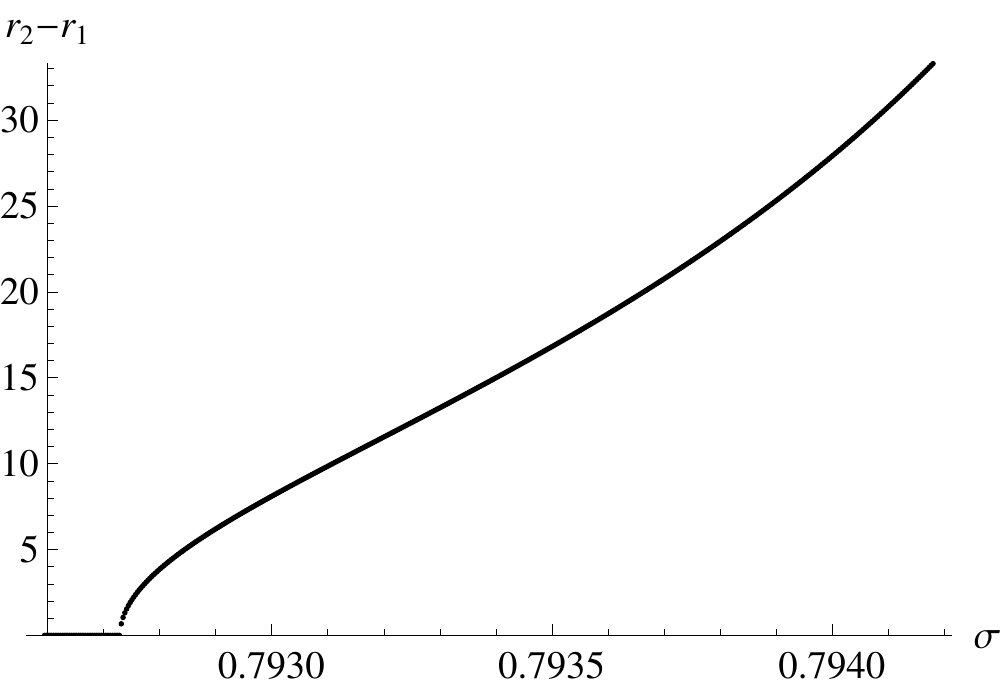}
\includegraphics[scale=0.7]{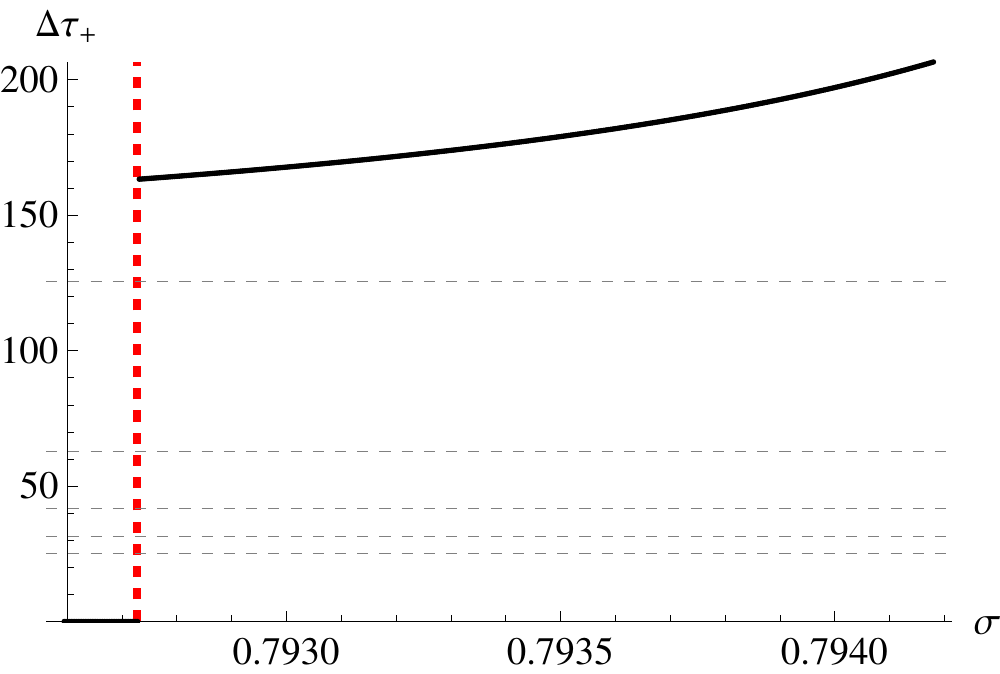}
\caption{\label{fig:r2mr1}Left: $r_{2} - r_{1}$ by varying $\sigma$, where we are considering $M = 5$ and $L_{-} = 0.1$. Right: $\Delta \tau_{+}$ as a function of $\sigma$. If $\sigma$ approaches $1/4\pi L_{-}$, then $\Delta \tau_{+}$ diverges. On the other side, if $\sigma$ approaches the degenerate limit, then it approaches the stationary shell limit (red dashed line), where we can identify with an arbitrary period, including $8\pi M/N$ (gray dashed lines are $8\pi M$, $8\pi M/2$, $8\pi M/3$, $8\pi M /4$, and $8\pi M/5$ from top to bottom).}
\end{center}
\end{figure}

We can show that for a given $M$ and $L_{-}$, there exists $\sigma$ that satisfies $\Delta \tau_{+} \times N = 8 \pi M$. We can rewrite the effective potential $V(r)$ as
\begin{eqnarray}
V(r) = 1 - \frac{M^{2}}{16 \pi^{2} \sigma^{2} r^{4}} - \frac{M}{16 \pi^{2} \sigma^{2}} \left(L_{-}^{-2} + 16 \pi^{2} \sigma^{2} \right) \frac{1}{r} - \frac{1}{64 \pi^{2} \sigma^{2}} \left(L_{-}^{-2} - 16 \pi^{2} \sigma^{2} \right)^{2} r^{2}.
\end{eqnarray}
First, if $4\pi \sigma L_{-} = 1$, then $V(r) = 1$ as $r$ goes to infinity. Therefore, $V(r)$ has only one zero and this corresponds to the limit when $\Delta \tau = \infty$. If $\sigma$ decreases infinitesimally from the limit $1 =4\pi \sigma L_{-}$ satisfying
\begin{eqnarray}
\frac{1}{L_{-}} \geq 4\pi \sigma,
\end{eqnarray}
then it allows two zeros. However, this is just a necessary condition. There may be a possibility that it allows no zeros with the condition $L_{-}^{-1} \geq 4\pi \sigma$. In order to find this limit, we think of the condition of $\sigma$ that satisfies a degenerate zero: $V(r_{0}) = V'(r_{0}) = 0$ with a zero $r_{0}$. In this case, the solution stops at a constant radius\footnote{Of course, we need to be careful that in this limit, the distance between $r_{2}$ and $r_{1}$ is the same order of the thickness of the shell, and hence, to describe this region more properly, we need to rely on what is beyond the thin-shell approximation. If we do this properly, then we expect that there should be a smooth transition from the two-zero limit to the degenerate limit.}, and hence one can identify an arbitrary period including $\Delta \tau = 0$. Note that the corresponding $r_{0}$ is
\begin{eqnarray}
r_{0}^{3} = M \left( \frac{\mathcal{C} + \sqrt{\mathcal{C}^{2} + 8}}{L_{-}^{-2} - 16 \pi^{2} \sigma^{2}} \right),
\end{eqnarray}
where
\begin{eqnarray}
\mathcal{C} \equiv \frac{L_{-}^{-2} + 16 \pi^{2} \sigma^{2}}{L_{-}^{-2} - 16 \pi^{2} \sigma^{2}}.
\end{eqnarray}
By plugging this $r_{0}$ into $V(r_{0}) = 0$, we can prove that there are two zeros if $M < M_{*}$, where
\begin{eqnarray}
M_{*} \equiv \frac{64 \pi^{3} \sigma^{3}}{\left( L_{-}^{-2} - 16 \pi^{2} \sigma^{2} \right)^{2}} \left[ 3 + \frac{3\mathcal{C}}{2} \left( \mathcal{C} + \sqrt{\mathcal{C}^{2} + 8} \right) \right]^{-3/2} \left( \mathcal{C} + \sqrt{\mathcal{C}^{2} + 8} \right)^{2}.
\end{eqnarray}
In FIG.~\ref{fig:range}, we plot $M_{*}$ as a function of $L_{-}$ and $4\pi \sigma$. For a given $L_{-}$, this $M_{*}$ ranges from zero to infinity, and so for a given $L_{-}$ and a given $M$, there is always a range of $\sigma$ that allows $M_{*} > M$. Therefore, for a given $L_{-}$ and $M$, we can smoothly scale $\Delta \tau$ from $0$ to $\infty$ by adjusting $\sigma$; and hence, (since $r > 2M$ and Eq.~(\ref{eq:redshi}) is a regular integration) there exists a proper $\sigma$ that satisfies $\Delta \tau_{+} \times N = 8 \pi M$ (FIG.~\ref{fig:r2mr1}). 

\subsection{Decay rates}

\subsubsection{Euclidean approach: Farhi-Guth-Guven/Gregory-Moss-Withers tunneling}

The decay rate is
\begin{eqnarray}
\Gamma \propto e^{-2B},
\end{eqnarray}
where
\begin{eqnarray}
B = S_{\mathrm{E}}(\mathrm{solution}) - S_{\mathrm{E}}(\mathrm{background})
\end{eqnarray}
and the Euclidean action is
\begin{eqnarray}
S_{\mathrm{E}} = - \int_{\mathcal{M}} \sqrt{+g} d^{4}x \left[ \frac{\mathcal{R}}{16 \pi} - \frac{1}{2} \nabla^{\mu} \phi \nabla_{\mu} \phi - U(\phi) \right] - \int_{\partial \mathcal{M}} \sqrt{+h} d^{3}x \left[ \frac{\mathcal{K} - \mathcal{K}_{0}}{8 \pi} \right].
\end{eqnarray}

If we want to interpret this as the nothing-to-something interpretation, in order to subtract the boundary terms at infinity, the Euclidean time of the background should be the same as that of the solution. This may make a cusp singularity of the solution part, but if the inside of the shell has zero mass, then we do not need to worry about the cusp singularity. Applying the thin-shell approximation, one can calculate the decay rate of the thin-shell bubbles. In the literature, there are independent but consistent derivations of the decay rate. Without derivation, we use the formula following Gregory-Moss-Withers \cite{Gregory:2013hja} (here, $'$ is a differentiation with respect to $r$; $\dot{}$ is a differentiation with respect to $\tau$),
\begin{eqnarray}\label{eq:GMW}
2B = \frac{\mathcal{A}_{\mathrm{i}} - \mathcal{A}_{\mathrm{f}}}{4} + \frac{1}{4} \int d\tau \left[ \left( 2rf_{+} - r^{2}f_{+}' \right) \dot{\tau}_{+} - \left( 2rf_{-} - r^{2}f_{-}' \right) \dot{\tau}_{-} \right],
\end{eqnarray}
where $\tau$ (Euclidean proper time of the shell) and $\tau_{\pm}$ (Euclidean time of the outside and inside geometry) satisfy
\begin{eqnarray}
f_{\pm}^{2} \dot{\tau}_{\pm}^{2} + \dot{r}^{2} &=& f_{\pm},\\
f_{\pm} \dot{\tau}_{\pm} &=& \beta_{\pm}.
\end{eqnarray}
Here, the first term of Eq.~(\ref{eq:GMW}) originates from the regularization of the cusp singularity.

By using this, one can present the second term as
\begin{eqnarray}
\frac{1}{4} \int dr \left[ \beta_{+} \frac{\left( 2r - r^{2}f_{+}'/f_{+} \right)}{\sqrt{f_{+} - \beta_{+}^{2}}} - \beta_{-} \frac{\left( 2r - r^{2}f_{-}'/f_{-} \right)}{\sqrt{f_{-} - \beta_{-}^{2}}}\right].
\end{eqnarray}
By using the identity $\beta_{+}^{2} - f_{+} = \beta_{-}^{2} - f_{-}$ and $\beta_{-}'r - \beta_{-} = \beta_{+}'r - \beta_{+}$, one can change the form
\begin{eqnarray}
\frac{1}{4} \int dr \left[ \frac{\left( 2 \beta_{+}' r^{2} - r^{2} \beta_{+} f_{+}'/f_{+} \right)}{\sqrt{f_{+} - \beta_{+}^{2}}} - \frac{\left( 2 \beta_{-}' r^{2} - r^{2} \beta_{-} f_{-}'/f_{-} \right)}{\sqrt{f_{-} - \beta_{-}^{2}}}\right].
\end{eqnarray}
Finally, by using the integration by parts, one can present this integration as equivalent to
\begin{eqnarray}
\int dr r \left[ \cos^{-1} \left(\frac{\beta_{+}}{\sqrt{f_{+}}}\right) - \cos^{-1} \left(\frac{\beta_{-}}{\sqrt{f_{-}}}\right) \right].
\end{eqnarray}
In addition, by using straightforward calculations, we can finally reach the following form \cite{Ansoldi:1997hz}:
\begin{eqnarray}
2B = \frac{\mathcal{A}_{\mathrm{i}} - \mathcal{A}_{\mathrm{f}}}{4} + 2 \int_{r_{1}}^{r_{2}} dr r \left| \cos^{-1} \left( \frac{f_{+} + f_{-} - 16 \pi^{2} \sigma^{2} r^{2}}{2\sqrt{f_{+}f_{-}}} \right) \right|.
\end{eqnarray}

\subsubsection{Hamiltonian approach: Fischler-Morgan-Polchinski tunneling}

According to Fischler-Morgan-Polchinski \cite{Fischler:1989se}, following the WKB approximation, the tunneling rate is
\begin{eqnarray}
\Gamma \propto e^{2 i \left(\Sigma_{\mathrm{f}} - \Sigma_{\mathrm{i}} \right)},
\end{eqnarray}
where the wave function is approximated by $\Psi \sim e^{i\Sigma}$ with a $\Sigma$ that satisfies the Hamilton-Jacobi equation and the metric ansatz is given by
\begin{eqnarray}
ds^{2} = - N^{t} dt^{2} + L^{2} \left( d\eta + N^{\eta} dt \right)^{2} + r^{2} d\Omega^{2},
\end{eqnarray}
where all metric functions $N^{t}$, $N^{\eta}$, $L$ and $r$ are functions of $\eta$ and $t$, where $\eta$ is defined over a space-like hypersurface. When we do the thin-shell approximation, on the inside or outside of the shell, the integration becomes
\begin{eqnarray}
i\Sigma_{\mathrm{vol}} = - \int_{\mathrm{vol}} d\eta \left[ r \sqrt{L^{2} f_{\pm} - r'^{2}} - r r' \cos^{-1}\left(\frac{r'}{L\sqrt{f_{\pm}}}\right) \right],
\end{eqnarray}
while $\eta$ covers inside ($-$) or outside ($+$) the shell (now $'$ is a differentiation with respect to $\eta$). The integration on the shell (between $\eta_{\mathrm{shell}} - \epsilon$ and $\eta_{\mathrm{shell}} + \epsilon$, where $\eta_{\mathrm{shell}}$ is the coordinate on the shell and $\epsilon$ is an arbitrary small number) becomes
\begin{eqnarray}
i\Sigma_{\mathrm{shell}} = \int^{r_{\mathrm{shell}}} dr \left[ r \cos^{-1}\left(\frac{r'(\eta_{\mathrm{shell}}-\epsilon)}{\hat{L}\sqrt{f_{-}}}\right) - r \cos^{-1}\left(\frac{r'(\eta_{\mathrm{shell}}+\epsilon)}{\hat{L}\sqrt{f_{+}}}\right) \right],
\end{eqnarray}
where the shell is on $r_{\mathrm{shell}}$. If the stationary shell condition is satisfied, then
\begin{eqnarray}
\frac{r'(\eta_{\mathrm{shell}}\pm\epsilon)}{\hat{L}} &=& \beta_{\pm},\\
\frac{r'}{L} &=& \sqrt{f_{\pm}}
\end{eqnarray}
for inside and outside the shell.

\begin{figure}
\begin{center}
\includegraphics[scale=0.75]{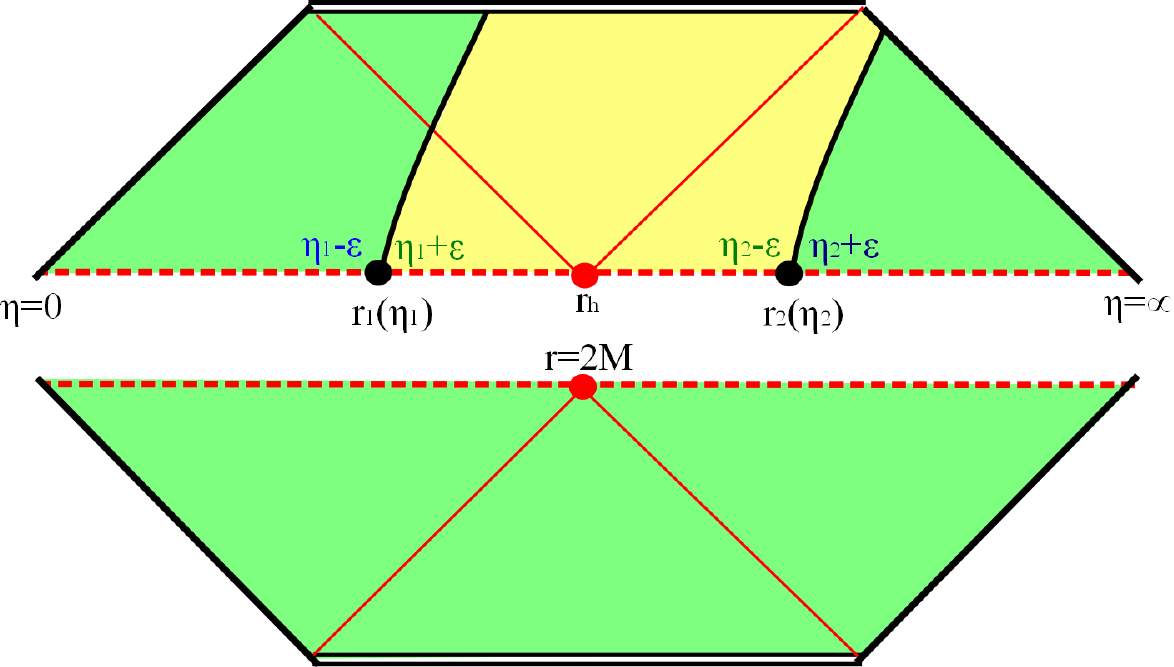}
\caption{\label{fig:FMP}Schematic picture for the Fischler-Morgan-Polchinski integration.}
\end{center}
\end{figure}

For the nothing-to-something interpretation, there is no shell initially and there are two shells after tunneling. On the final hypersurface, there are two shells. Therefore, the integral is presented as follows (FIG.~\ref{fig:FMP}):
\begin{eqnarray}
\int_{0}^{\eta_{1}-\epsilon} d\eta \left( ... \right) + \int_{\eta_{1}-\epsilon}^{\eta_{1}+\epsilon} d\eta \left( ... \right) + \int_{\eta_{1}+\epsilon}^{\eta_{2}-\epsilon} d\eta \left( ... \right) + \int_{\eta_{2}-\epsilon}^{\eta_{2}+\epsilon} d\eta \left( ... \right) + \int_{\eta_{2}+\epsilon}^{\infty} d\eta \left( ... \right), 
\end{eqnarray}
where $\eta_{1}$ is the position of the left shell and $\eta_{2}$ is the position of the right shell.

\paragraph{Shell integration} The second and fourth terms are the integration over the shell. Note that $\eta_{1}-\epsilon$ and $\eta_{2}+\epsilon$ are the same outside geometry while $\eta_{2}-\epsilon$ and $\eta_{1}+\epsilon$ are the same inside geometry. Therefore, if we change the integration as follows
\begin{eqnarray}
-\int_{\eta_{1}+\epsilon}^{\eta_{1}-\epsilon} + \int_{\eta_{2}-\epsilon}^{\eta_{2}+\epsilon} \left( ... \right),
\end{eqnarray}
then two integrals share the \textit{common} integrand. Now by changing the variable to $r$ integration, the first integration is to $r_{1}$ (left shell) and the second integration is to $r_{2}$ (right shell); then we can present this integration by
\begin{eqnarray}
\int_{r_{1}}^{r_{2}} dr \left( ... \right).
\end{eqnarray}
Therefore, one can easily prove that this gives the same result of the second term of the Gregory-Moss-Withers tunneling \cite{Gregory:2013hja}
\footnote{If $\Delta \tau_{+} \times N = 8 \pi M$ with even number $N$, then the thin-shell integration can vanish since $r_{1} = r_{2}$; like this, if $N$ is an odd number greater than $3$, then the Hamiltonian approach can underestimate the shell integration. In these cases, we need to follow the Euclidean approach rather than the naive results of the Hamiltonian approach. One can interpret that the Hamiltonian approach (WKB approximation) considers the most probable history of a tunneling process, while the Euclidean approach can cover more various solutions that cannot be covered by the WKB approximation. There is an interesting analogy with oscillating instantons in $O(4)$-symmetric cases \cite{Hackworth:2004xb}. In addition, relations with the negative modes could be a future interesting topic \cite{Battarra:2012vu}.}.

\paragraph{Volume integration} There remain volume integrations:
\begin{eqnarray}
\int_{0}^{\eta_{1}} d\eta \left( ... \right) + \int_{\eta_{1}}^{\eta_{2}} d\eta \left( ... \right) + \int_{\eta_{2}}^{\infty} d\eta \left( ... \right),
\end{eqnarray}
where the second integration is over the inside geometry while the first and third integrations are over the outside geometry. Note that because of the stationary shell condition ($r' = L\sqrt{f_{\pm}}$), the only contribution comes from the $\arccos$ integration, where the $\arccos$ term is $\pi$ if $r' < 0$ (beyond the Einstein-Rosen bridge) or $0$ if $r' > 0$. In the end, the volume term contributes
\begin{eqnarray}
i\Sigma_{\mathrm{f, vol}} = -\frac{\pi}{2} \left(r_{\infty}^{2} - r_{1}^{2} \right) - \frac{\pi}{2} \left(r_{1}^{2} - r_{h}^{2} \right),
\end{eqnarray}
where $r_{h}$ is the horizon radius of the internal geometry and $r_{\infty} = \infty$. This should be subtracted by the initial hypersurface integration:
\begin{eqnarray}
i\Sigma_{\mathrm{i}} = -\frac{\pi}{2} \left(r_{\infty}^{2} - r_{+}^{2} \right),
\end{eqnarray}
where $r_{+} = 2M$ is the initial horizon radius. Finally, the subtracted volume term becomes
\begin{eqnarray}
-2i \left( \Sigma_{\mathrm{f, vol}} - \Sigma_{\mathrm{i}} \right) = \pi \left(r_{+}^{2} - r_{h}^{2} \right) = \frac{\mathcal{A}_{\mathrm{i}}-\mathcal{A}_{\mathrm{f}}}{4}.
\end{eqnarray}
This gives the first term of the Gregory-Moss-Withers tunneling \cite{Gregory:2013hja}.

\paragraph{\label{sec:comm}Comments on cusp singularities}

We focused on the case when the internal geometry is anti de Sitter, i.e., $M_{-} = 0$. If $M_{-} > 0$, then in general there appears a cusp singularity at the horizon of the internal geometry. This can be regularized by a certain scheme as Gregory-Moss-Withers did \cite{Gregory:2013hja}. This regularization scheme could be doubted since it is a kind of singularity. On the other hand, if this regularized result can be justified by another independent way, then we can trust the regularization method. Note that the volume terms of Fischler-Morgan-Polchinski \cite{Fischler:1989se} exactly give the regularization terms of Gregory-Moss-Withers \cite{Gregory:2013hja}. This can be an independent justification of the regularization technique as well as the resolution of subtleties of the instanton approach (as was observed by \cite{Ansoldi:2014hta}).

\subsection{Applications to the information loss problem}

The nothing-to-something interpretation can be understood as the decay of a black hole. This can be especially applied to the information loss problem \cite{Hawking:1976ra}.

If $M_{-} > 0$, then this implies that a quantum fluctuation that emits a large mass (although the probability is exponentially suppressed) can cause a bias from the adiabatic process \cite{Chen:2015gux}. Therefore, the true event horizon $r_{\mathrm{EH}}$ can be different from that of the putative event horizon ${r'}_{\mathrm{EH}}$, where this would be the event horizon if there are only adiabatic processes. If $|r_{\mathrm{EH}} - {r'}_{\mathrm{EH}}| \propto M_{+} - M_{-} \gg \ell_{\mathrm{Pl}}$ and the firewall \cite{Almheiri:2012rt} could be assumed to grow around ${r'}_{\mathrm{EH}}$, then it can be a good proof that the firewall becomes observable from a distance, due to this non-adiabatic fluctuating process \cite{Chen:2015gux}.

If $M_{-} = 0$, then this instanton induces a trivial geometry. In the Euclidean path-integral approach, the propagator between the initial hypersurface and final hypersurface can be presented by
\begin{eqnarray}
\langle f | i \rangle = \int_{i\rightarrow f} \mathcal{D} g \mathcal{D} \phi \; e^{- S_{\mathrm{E}}} \simeq \sum_{i \rightarrow f} e^{- S_{\mathrm{E}}^{\mathrm{ins}}},
\end{eqnarray}
where we sum over all metrics and fields that connect hypersurfaces $i$ and $f$; in the last part of the above equation, this path-integral can be well approximated by a sum over on-shell histories (instantons). Among the instanton paths that connect from $i$ to $f$, if there is a trivial geometry without horizons nor singularities, e.g., the periodically identified anti-de Sitter space, then it is used to recover correlations in the end, as was emphasized by \cite{Maldacena:2001kr,Hawking:2005kf}. Therefore, as long as there exists such an instanton with $M_{-} = 0$, it will help recover correlations and will be well embedded in the scenario of the effective loss of information \cite{Sasaki:2014spa}.

\section{\label{sec:dis}Discussion}

In this paper, we focused on two complementary interpretations of instantons. There are two types of interpretations, what we named the nothing-to-something interpretation, when the initial surface and final surface are disconnected by Euclidean geometries, and what we named the something-to-something interpretation, when the initial surface and final surface are connected by a Euclidean geometry. For Coleman-DeLuccia instantons \cite{Coleman:1980aw}, the nothing-to-something interpretation is rather usual, while for thin-shell instantons \cite{Farhi:1989yr}, the something-to-something interpretation is usual. On the other hand, a rather unusual interpretation is possible not only for Coleman-DeLuccia instantons \cite{Brown:2007sd}, but also for thin-shell instantons with some restricted conditions. We obtained a consistent decay rate by both approaches: the Euclidean approach \cite{Farhi:1989yr,Gregory:2013hja} and Hamiltonian approach \cite{Fischler:1989se}. One important comment is that the two approaches are not coincide with each other, if the boundary term at infinity is not canceled; hence, if $\Delta \tau_{+} = 8 \pi M$ is not satisfied, then we cannot do a consistent interpretation.

It is interesting that two independent approaches coincide with each other. The volume term of the Hamiltonian approach \cite{Fischler:1989se} corresponds to the regularization term around the cusp singularity of the Euclidean geometry \cite{Gregory:2013hja}. This shows that the regularization technique of the Euclidean manifold is indeed in a right way. This justification helps us to investigate more general instantons, where we remain for possible future projects.

For a thin-shell instanton, if the nothing-to-something interpretation is possible, then we can further interpret that a stationary black hole decays and emits an out-going shell; and finally a black hole decreases its mass or even disappears. One remark is that FIG.~\ref{fig:ads_single} is related to the work of Hartle and Hawking \cite{Hartle:1976tp}. In this path-integral derivation of Hawking radiation \cite{Hartle:1976tp}, they constructed a tunneling of a particle (energy $\omega \ll M$, and hence one may neglect the back-reaction due to the emission of the particle), where the particle moves from inside to outside the black hole, and first moves backward in time and second moves forward in time. This process is not allowed classically, but the entire wave function allows such a process; and the entire wave function can be approximated by a classical path that is analytically continued by the Euclidean time. Our result FIG.~\ref{fig:ads_single} can be interpreted as a generalization of \cite{Hartle:1976tp}, but in our case, we can even consider the case that the emitted energy is comparable with the original black hole mass, since we have considered the back-reaction precisely.

Since a black hole can disappear by a quantum process, this can shed some light on the information loss problem \cite{Hawking:1976ra}. This is not yet a very general solution, but if at once such a process exists, then information can be conserved through such a process \cite{Maldacena:2001kr,Hawking:2005kf,Sasaki:2014spa}. In terms of the entire wave function, information should be conserved but the classical equations of motion including general relativity may not need to be satisfied due to the superposition of classical geometries, and hence this can be interpreted as the firewall phenomena \cite{Almheiri:2012rt}, while there is no real firewall that explicitly violates general relativity within a semi-classical background that can even be naked \cite{Hwang:2012nn,Chen:2015gux}. On the other hand, for a semi-classical geometry, it satisfies local quantum field theory and general relativity, while it violates unitarity; since Hawking radiation does not contain information, it can be free from troubles with black hole complementarity \cite{Susskind:1993if,Yeom:2008qw}. In this sense, we may name this idea the effective loss of information, since information is lost by a semi-classical observer, while the entire wave function conserves information, although it is fair to say that we need to generalize more on this process.

\newpage

\section*{Acknowledgment}

DY would like to thank Erick Weinberg, Bum-Hoon Lee, and Wonwoo Lee for stimulating discussions on the thermal interpretation of Coleman-DeLuccia instantons. PC is supported in part by the National Center for Theoretical Sciences (NCTS) and Ministry of Science and Technology (MOST) of Taiwan. DY is supported by Leung Center for Cosmology and Particle Astrophysics (LeCosPA) of National Taiwan University (103R4000).

\end{document}